\documentclass[
  reprint,
  superscriptaddress,
  amsmath,amssymb,
  aps,pra,
  floatfix
]{revtex4-2}

\usepackage[english]{babel}
\usepackage{graphicx}% Include figure files
\usepackage{dcolumn}% Align table columns on decimal point
\usepackage{bm}% bold math
\usepackage[
  colorlinks=true,
  linkcolor=blue,
  citecolor=blue,
  urlcolor=blue
]{hyperref}

\begin{document}

\preprint{APS/PhysRevA}

\title{Enhanced nondipole momentum offsets in  triple ionization of atoms driven by mid-infrared laser fields}% Force line breaks with \\

\author{S. J. Praill}

\author{G. P. Katsoulis}

\author{D. Romero Torres}

\author{A. Emmanouilidou}

\affiliation{%
Department of Physics and Astronomy, University College London,
Gower Street, London WC1E 6BT, United Kingdom
}

\date{\today}

\begin{abstract}
We investigate the dependence on wavelength of nondipole effects in triple ionization of Ne driven by intense infrared and mid-infrared laser pulses using a three-dimensional semiclassical model that fully accounts for nondipole effects and the Coulomb singularity in the electron-core interaction. We find in triple ionization  of strongly driven Ne a large positive average momentum offset along the direction of laser propagation, which vanishes in the dipole approximation. This positive momentum offset significantly increases with increasing wavelength of the laser pulse. This increase is present for all triple ionization events as well as for the main direct and delayed pathways of triple ionization. We attribute the increase of the momentum offset  to the contribution of the effect of the magnetic field of the laser pulse on the bound electrons. This contribution counterbalances the decrease of electron-electron correlation in recollisions with increasing wavelength. We find that  1200 nm  is  an appropriate wavelength  for experimentally measuring the momentum offset related to  correlated three-electron escape.

\end{abstract}

\maketitle

%\tableofcontents

\section{\label{sec:Introduction}Introduction}

Atoms and molecules driven by intense infrared laser fields exhibit a wide variety of interesting phenomena. One prominent example is nonsequential multiple ionization (NSMI) in atoms, a fundamental process mediated by strong electron–electron correlation \cite{Corkum_1994}. A fully \textit{ab initio} description of NSMI remains out of reach. Theoretical approaches  employ  approximations to reduce the complexity of the problem \cite{PhysRevLett.97.083001,Zhou:10,Tang:13,PhysRevA.98.031401,Prauzner_Bechcicki_2021,Efimov:21,PhysRevA.105.053119}. For instance, such an approximation is the dipole one. It neglects the spatial dependence of the vector potential $\mathbf{A}(\mathbf{r},t)$ and as a result  of the magnetic field of the laser pulse $\mathbf{B}(\mathbf{r},t)=\nabla\times\mathbf{A}(\mathbf{r},t)$. Nondipole effects for single ionization are expected to be important when the amplitude of the electron motion due to the Lorentz force, $\beta_0 \approx U_p / (2 \omega c)$, is approximately equal to one  \cite{Magnetic1,Magnetic2};  $U_p$ is the average energy an electron gains in the field,  known as the ponderomotive energy.  

Nondipole effects were identified in stabilization \cite{stabilization}, high-harmonic generation (HHG) \cite{HHG1,HHG2,HHG3}, multi-electron escape \cite{Neon} and many other processes in strongly driven systems \cite{PhysRevLett.106.193002,KellerMagnetic2014,CorkumBandrauk2015,Biegert2015,Emmanouilidou1,Emmanouilidou2,KellerMagneticFields,FSun,PhysRevLett.128.113201,Xe_nondipole,BurgdorferPRL2022}. In single ionization, the average final electron momentum along the direction of light propagation, which is zero in the dipole approximation, was observed experimentally \cite{PhysRevLett.106.193002} and computed \cite{CorkumBandrauk2015} to be positive. For HHG, it was shown that the tunnel-ionizing electron returns to the core only if it has an initial momentum that compensates for the drift due to the magnetic field \cite{HHG2}, or if a second laser field compensates for this drift \cite{HHG1}. For a review on nondipole effects in atomic ionization also see Ref.~\cite{Becker2012RMP}. For multi-electron escape, a three-dimensional semiclassical rescattering model that accounts for the magnetic field was used to describe nondipole effects in  the observed ionization of Ne$^{n+}$ ($n\leq 8$) in ultrastrong fields \cite{Neon}. 
 
We previously identified nondipole gated ionization, a nondipole effect in nonsequential double ionization (NSDI) in He \cite{Emmanouilidou1,Emmanouilidou2}. We did so at intensities that are orders of magnitude smaller than previously expected \cite{Magnetic1,Magnetic2}. That is, we found that correlated electron dynamics probes nondipole effects at intensities significantly smaller than for single ionization \cite{Emmanouilidou1,Emmanouilidou2}. Our calculations in Refs. \cite{Emmanouilidou1,Emmanouilidou2} were performed using a predecessor of the three-dimensional semiclassical model we employ in this work. In nondipole gated ionization the magnetic field along with the recollision act as a gate.  A recollision occurs when the tunnel-ionizing electron returns to recollide with the core \cite{Corkum_1994}. The resulting gate allows ionization to occur only for a subset of the initial momenta along the direction of light propagation of the recolliding electron  \cite{Emmanouilidou2}. This gate gives rise to a  negative shift of the average initial electron momentum. This negative shift compensates for the positive momentum shift induced by the Lorentz force along the direction of light propagation. Hence, this negative shift allows  for the electron to return to the core and recollide.
  In contrast, in the dipole approximation, the initial  momentum of the recolliding electron averages to zero. Also, for double ionization in He, we  found that the average final momentum of the recolliding electron, as well as the average sum of the final electron momenta along the direction of light propagation ($y$ axis)  has a significant positive offset. 
  These nondipole effects have been experimentally verified for driven Ar \cite{FSun}. Nondipole effects in double ionization have also been measured experimentally in driven Xe \cite{PhysRevLett.128.113201}. 
  
We also identified nondipole effects in nonsequential triple ionization (NSTI) of Ne  \cite{katsoulis_nondipole_2023}. We did so  by employing a three-dimensional semiclassical model that addresses artificial autoionization \cite{Agapi3electron, AgapiNeonPRL}. Classically, a bound electron can acquire very large energy.  This energy can be transferred to another bound electron via Coulomb interaction leading to its artificial escape. Our model addresses artificial autoionization  by employing effective Coulomb potentials to describe electron-electron interaction between two bound electrons (ECBB model) \cite{Agapi3electron,AgapiNeonPRL}. In the ECBB model, all other forces are fully accounted for.  That is, the Coulomb interaction of each electron with the nucleus and of each bound electron with a quasifree electron and  the interactions of all particles with the laser field are treated with no approximation.
 
 In NSTI of Ne, we previously found that nondipole gated ionization and  a significant positive offset of the average sum of the final electron momenta along the $y$ axis are also present \cite{katsoulis_nondipole_2023}, as for NSDI of He. However, at 800 nm wavelength of the laser pulse, the mechanism underlying the positive offset for the strong recollisions in Ne differs  from the respective one   for the glancing recollisions in  He.  For NSTI in Ne, we previously found that the change in momentum due to the magnetic field plays an important role in the positive momentum offset along the $y$ axis, which is not the case for NSDI in He. Also, for Ne, the magnitude of the positive momentum offset has a one-to-one correspondence with the degree of electron-electron  correlation in the recollisions resulting in multi-electron escape. Hence, we identified this positive momentum offset as a probe of the strength of electronic correlation in NSTI in Ne  \cite{katsoulis_nondipole_2023}.
 
Here, we identify the mid-infrared wavelength of the laser field for which  the nondipole effects we previously identified in strongly driven Ne \cite{katsoulis_nondipole_2023} become significantly enhanced while at the same time nonsequential triple ionization occurs with probability comparable to the one at 800 nm.  In the current study, we employ the extended ECBB model \cite{praill_three-electron_2026} that involves  a generalized  initial state of the tunnel-ionizing electron and also accounts for electrons  tunnelling during time propagation. 

We consider triple ionization of Ne driven by  a mid-infrared laser pulse at 1200 nm and at 1600 nm and compare with triple ionization of Ne driven by a laser pulse at 800 nm, the latter considered in Ref. \cite{katsoulis_nondipole_2023}. We do the same for double ionization of strongly driven Ne. However, for DI, we only compare our results at 1200 nm with the ones at 800 nm, since we find that at 1600 nm for half of the DI events no recollision occurs. 
 We find that the positive final  electron-momentum shift is significantly enhanced when Ne is driven by a pulse at 1200 nm and at 1600 nm compared to  800 nm; all pulses have the same intensity of 1.6 PW/cm$^2$.  This enhancement is present for nonsequential triple ionization (NSTI) and nonsequential double ionization (NSDI).  However, the probability for nonsequential triple ionization is significantly larger at 1200 nm compared to 1600 nm. Hence, the current study suggests that 1200 nm is an appropriate wavelength for experiments to measure nondipole effects in triple ionization of strongly driven Ne.

\section{\label{sec:Method}Method}

We employ the ECBB model~\cite{Agapi3electron, AgapiNeonPRL}  to compute triple ionization (TI) and double ionization (DI) observables for strongly driven Ne. Specifically, we employ the extended ECBB model~\cite{praill_three-electron_2026}, where we generalize the initial state of the tunnel-ionizing electron  and we allow for the electrons to tunnel during time propagation~\cite{praill_three-electron_2026}. We employ the extended ECBB model, since for Ne driven by a pulse at 1200 nm and  1600 nm we find that tunnelling during time propagation is important. Namely, we find that double ionization   is significantly suppressed when tunnelling is turned off compared to when tunnelling is turned on  during time propagation.

The ECBB model addresses the challenge facing classical and semiclassical models of NSMI, namely artificial  autoionization. It does so by employing effective Coulomb potentials to treat the interaction between any pair of bound electrons. To address artificial autoionization, some classical models soften the Coulomb potential~\cite{PhysRevLett.97.083001,Zhou:10,Tang:13}.
Other semiclassical methods  employ Heisenberg potentials that mimic the Heisenberg uncertainty principle,  preventing a bound electron from undergoing a close encounter with the core~\cite{PhysRevA.21.834}. Effectively, these latter models also soften the Coulomb  interaction of an electron with the core. Both of these approaches fail to accurately describe electron scattering from the core~\cite{Pandit2017,Pandit2018}, which underlies nonsequential multiple ionization.  The ECBB model addresses  artificial autoionization, while fully accounting for the electron-core interaction, resulting in an accurate treatment of NSMI.

In the ECBB model, one electron tunnels through the field lowered Coulomb barrier at time $t_0$. As in our previous studies of TI~\cite{Agapi3electron,katsoulis_nondipole_2023,AgapiNeonPRL,praill_three-electron_2026}, one electron (electron 1) tunnel-ionizes with a rate described by the quantum mechanical Ammosov-Delone-Krainov (ADK) formula~\cite{Delone:91,Landau} with empirical corrections for high intensities~\cite{Tong_2005}. This rate is also adjusted to account for the depletion of the initial ground state~\cite{AgapiNeonPRL}. The time   electron 1 tunnels, $t_0$, is obtained randomly 
in the time interval where the electric field is nonzero, using importance sampling~\cite{ROTA1986123}. We take the time interval to be  [-2$\tau$, 2$\tau$], where $\tau$ is the full width at half maximum of the pulse duration in intensity. The exit point of electron 1 along the direction of the electric field is obtained either analytically using parabolic coordinates or numerically assuming the electron exits along the direction of the laser field,  see Ref.~\cite{praill_three-electron_2026}. For the parameters of the laser field considered here, we find that the exit point of  electron 1 is essentially determined by parabolic coordinates. We set equal to zero the momentum of electron 1 along the direction of the electric field, while  the transverse one is given by a Gaussian distribution~\cite{Delone:91,Delone_1998,PhysRevLett.112.213001}. This Gaussian distribution is centered around zero, since nondipole effects are not accounted for in the tunnelling rate. When nondipole effects are accounted for during tunnelling, the most probable momentum of the tunnel ionizing electron along the direction of light propagation is at most $I_{p,1}/3c$ \cite{Klaiber2013,keitel,Eckey2023}, with $I_{p,1}$ the first ionization potential. For Ne, $I_{p,1}/3c$ is equal to $1.9\times10^{-3}$ a.u. Since this value is significantly smaller than the electron momentum offsets presented in what follows, we are justified in neglecting it. For the remaining bound electrons, we employ a microcanonical distribution~\cite{Agapi3electron}.

The ECBB model is developed in a nondipole framework that  fully accounts for the magnetic field of the laser pulse, i.e., the magnetic field component of the Lorentz force. We use a vector potential of the form
\begin{equation}\label{eqn::vector_potential}
    \mathbf{A}(y, t)=-\frac{E_0}{\omega} \exp \left[-2 \ln 2\left(\frac{c t-y}{c \tau}\right)^2\right] \sin (\omega t-k y) \hat{\mathbf{z}},
\end{equation}
where $k=\omega/c$ is the wave number of the laser field. The direction of the vector potential and the electric field, $\mathbf{E}(y,t)=-\frac{\partial \mathbf{A}(y,t)}{\partial t}$, is along the $z$ axis, while the direction of light propagation is along the $y$ axis. The magnetic field, $\mathbf{B}(y,t)=\nabla \times \mathbf{A}(y,t)$, points along the $x$ axis. 

The Hamiltonian of the four-body system is given by
\begin{equation}\label{eqn::hamiltonian}
\begin{aligned}
H ={}&
\sum_{i=1}^{4}
\frac{\left[\tilde{\mathbf{p}}_{i}- Q_i \mathbf{A}(y_i,t)\right]^2}{2m_i}
+ \sum_{i=1}^{3}
\frac{Q_i Q_4}{\lvert \mathbf{r}_i-\mathbf{r}_4\rvert} 
\\[6pt]
&+
\sum_{i=1}^{2}\sum_{j=i+1}^{3}
\left[1-c_{i,j}(t)\right]
\frac{Q_i Q_j}{\lvert \mathbf{r}_i-\mathbf{r}_j\rvert}
\\[6pt]
&+
\sum_{i=1}^{2}\sum_{j=i+1}^{3}
c_{i,j}(t)
\Big[
V_{\mathrm{eff}}\!\left(\zeta_j(t),\lvert \mathbf{r}_4-\mathbf{r}_i\rvert\right) \\
& \qquad \qquad \qquad \quad
+
V_{\mathrm{eff}}\!\left(\zeta_i(t),\lvert \mathbf{r}_4-\mathbf{r}_j\rvert\right)
\Big].
\end{aligned}
\end{equation}
The function $c_{i,j}(t)$ determines whether the interaction between electrons $i$ and $j$ is described by the full Coulomb potential or by the effective Coulomb potentials. Specifically, $c_{i,j}(t)=0$ when at least one electron of the pair is quasifree, in which case the full Coulomb interaction is employed, while $c_{i,j}(t)=1$ when both electrons are bound, in which case the effective Coulomb potentials are employed. To ensure a smooth transition between the full and the effective Coulomb potential, $c_{i,j}(t)$ changes linearly between zero and one when either electron changes between bound and quasifree. Further details on the definition of $c_{i,j}(t)$ can be found in Ref.~\cite{Agapi3electron}. In Eq.~\eqref{eqn::hamiltonian}, $Q_i$ and $m_i$ are the charge and mass, respectively, while $\mathbf{r}_i$ and $\tilde{\mathbf{p}}_i$ are the position and canonical momentum vectors of particle $i$. The mechanical momentum $\mathbf{p}_i$ is given by
\begin{equation}
\mathbf{p}_{i}=\mathbf{\tilde{p}}_{i}-Q_i\mathbf{A}(y_i,t).
\end{equation}
The effective Coulomb potential felt by an electron $i$, at a distance $|\mathbf{r}_4-\mathbf{r}_i|$ from the core, which is labelled as particle 4 with $Q_4$ = 3 a.u., due to the charge distribution of electron $j$ is equal to
\begin{equation} \label{eqn::effective_potential}
    V_{\mathrm{eff}}(\zeta_j,|\mathbf{r}_{4}-\mathbf{r}_{i}|) =  \dfrac{1-(1+\zeta_j| \mathbf{r}_{4}-\mathbf{r}_{i}|)e^{-2\zeta_j| \mathbf{r}_{4}-\mathbf{r}_{i}|}}{| \mathbf{r}_{4}-\mathbf{r}_{i}|}. 
\end{equation}
We utilize this effective potential to prevent artificial autoionization. As electron $i$ approaches the core, i.e., $\mathbf{r}_i \to \mathbf{r}_4$,  $V_{\mathrm{eff}}(\zeta_j,|\mathbf{r}_4-\mathbf{r}_i|) \to \zeta_j$. This places a limit on the energy transfer between electrons $i$ and $j$, ensuring that artificial autoionization does not occur. This potential is turned on at the start and during time propagation only between pairs of electrons that are identified as bound. We note that we denote by 2 and 3 the initially bound electrons.

The effective charge $\zeta_{j}(t)$ is expressed in terms of the energy $\mathcal{E}_{j}(t)$ of electron $j$ as follows
\begin{equation}
\zeta_j(t)=
\begin{cases}
Q_4, & \mathcal{E}_j(t)\leq\mathcal{E}_{\mathrm{1s}},\\[2mm]
-\dfrac{Q_4}{\mathcal{E}_{\mathrm{1s}}}\mathcal{E}_j(t),
& \mathcal{E}_{\mathrm{1s}}<\mathcal{E}_j(t)<0,\\[2mm]
0, & \mathcal{E}_j(t)\geq0.
\end{cases}
\label{eq:zeta_definition}
\end{equation}
In Eq. \eqref{eq:zeta_definition}, $\mathcal{E}_{\mathrm{1s}}$ is the ground-state energy of a hydrogenic atom, i.e., $\mathcal{E}_{\mathrm{1s}} = - \; Q_4^2/2.$  The energy $\mathcal{E}_j(t)$ of electron $j$ is given by
{\allowdisplaybreaks
\begin{align}\label{eq:energy_of_electron_j}
&\mathcal{E}_{j}(t) = \frac{\left[\tilde{\mathbf{p}}_{j}- Q_j\mathbf{A}(y_j,t) \right]^2}{2m_j} + \frac{Q_4 Q_j}{|\mathbf{r}_{4}-\mathbf{r}_{j}|} \nonumber \\
%%%%%%%%%%%%%%%%%%%%%%%%%%%%%%%%%%%%%%
&+ \sum_{\substack{\;{i=1} \\ {i} \neq {j}}}^{3} c_{i,j}(t) V_{\mathrm{eff}}(\zeta_{i}(t),|\mathbf{r}_{4}-\mathbf{r}_{j}|) - Q_j\mathbf{r}_{j} \cdot \mathbf{E}\left(y_j, t\right).
\end{align}}
Hence, we compute the effective charge, $\zeta_{j}(t)$, on the fly, i.e.,  $\zeta_{j}(t)$ is changing as a function of time. At the start of propagation, i.e., at time $t_0$ we compute the effective charges of the bound electrons by setting the energy $\mathcal{E}_{j}(t)$ equal to the second ionization potential, $I_{p,2}.$ 

A sophisticated aspect of the ECBB model is that we determine on the fly whether an electron is quasifree or bound using a set of simple criteria. These criteria are discussed in detail and illustrated in~\cite{AgapiNeonPRL,Agapi3electron}. Briefly, according to these criteria, following a recollision with the core, a quasifree electron  transitions to bound  when its dynamics along the $z$ axis becomes influenced more by the core than the electric field. A bound electron transitions to quasifree due to a transfer of energy either from a recollision or from the laser field. More details on how a recollision between electrons is identified can be found in Appendix~\ref{Appendix_A}.

In our formulation, we fully account for the Coulomb singularities. Hence, an electron can approach infinitely close to the nucleus during time propagation. To ensure the accurate numerical treatment of the N-body problem in the laser field, we perform a global regularization. This regularization was introduced in the context of the gravitational N-body problem~\cite{Heggie1974}. Here, we integrate the equations of motion using a leapfrog technique~\cite{Pihajoki2015,Liu2016} jointly with the Bulirsch-Stoer method~\cite{press2007numerical}; see Ref.~\cite{PhysRevA.103.033115} for further details. We stop the propagation of each trajectory once the energy of each particle has converged. We label a trajectory as triple ionized if all three electrons at the end of the time propagation have positive energy.

\section{\label{sec:Results}Results}

Using the extended ECBB model, we study and compare nondipole effects in  TI of Ne driven by an intense laser pulse at 1600 nm, 1200 nm and 800 nm and in DI of Ne driven by an intense laser pulse at 1200 nm and 800 nm. All pulses correspond to roughly nine optical cycles, corresponding to a duration of 50 fs at 1600 nm, 37.5 fs at 1200 nm, and 25 fs at 800 nm. We consider  the same intensity of 1.6 PW/cm$^2$ for all pulses.  In what follows, we show that nondipole effects in TI  are enhanced when Ne is driven by a mid-infrared laser pulse at 1200 nm and 1600 nm compared to when Ne is driven by a pulse at 800 nm. We show the same for DI when Ne is driven by a laser pulse at 1200 nm compared to when Ne is driven by a pulse at 800 nm.  Also, we find that the probability for TI of Ne when driven by a   laser pulse at 1600 nm is more than an order of magnitude smaller compared to 1.1$\times$ 10$^{-2}$ \% at 800 nm. However, the probability for  TI of Ne when driven by a laser pulse at 1200 nm is half the respective probability at 800 nm. Hence, Ne driven by a laser pulse at 1200 nm is most appropriate to measure  experimentally  large nondipole effects in TI of Ne comparable to the nondipole effects in DI of He~\cite{Emmanouilidou1,Emmanouilidou2}, which have already been accessed experimentally. 
%In the following figures, we express momenta in units of $\sqrt{U_p}$, where $U_p = E_0^2/(4 \omega ^2)$. 

For Ne, to our knowledge, there are no TI experimental results to compare with for  mid-IR laser pulses. However, there are DI experimental results for Ne driven at 800 nm as well as at 1300 nm. For DI of Ne driven at 800 nm, comparing the sum of the momenta of the escaping electrons along the polarization direction with experimental results at 0.8 $\mathrm{PW/cm^2}$,  1 $\mathrm{PW/cm^2}$~\cite{deJesus_2004} and at 1.3 $\mathrm{PW/cm^2}$~\cite{PhysRevLett.84.447}, we find that the agreement for DI is not as good as  for TI, which we demonstrated in previous work~\cite{AgapiNeonPRL}. This comes as no surprise, since in the ECBB model we expect to have better agreement when  all active electrons escape (TI process here), since the influence of the approximate effective potentials is less for TI where all active electrons escape. Also, for DI of Ne driven at 1300 nm, comparing the sum of the momenta of the escaping electrons along the polarization direction  at 0.4 $\mathrm{PW/cm^2}$~\cite{Herrwerth_2008}, 0.5  $\mathrm{PW/cm^2}$ and 1.2  $\mathrm{PW/cm^2}$~\cite{Alnaser_2008} with experimental results, we find that the agreement with experiment is better at 1300 nm than at 800 nm.  Given that the agreement of our DI results is better for longer wavelength pulses and that our TI results at 800 nm  have excellent agreement with experiment, we expect that the agreement of our  TI results for mid-infrared pulses will be very good with future experimental results.

\subsection{Positive momentum offset of TI and DI in strongly driven Ne as a function of the wavelength of the laser pulse}\label{Section:Positive_mom_offset}

\begin{figure}[b]
	\centering
\includegraphics[width=\linewidth]{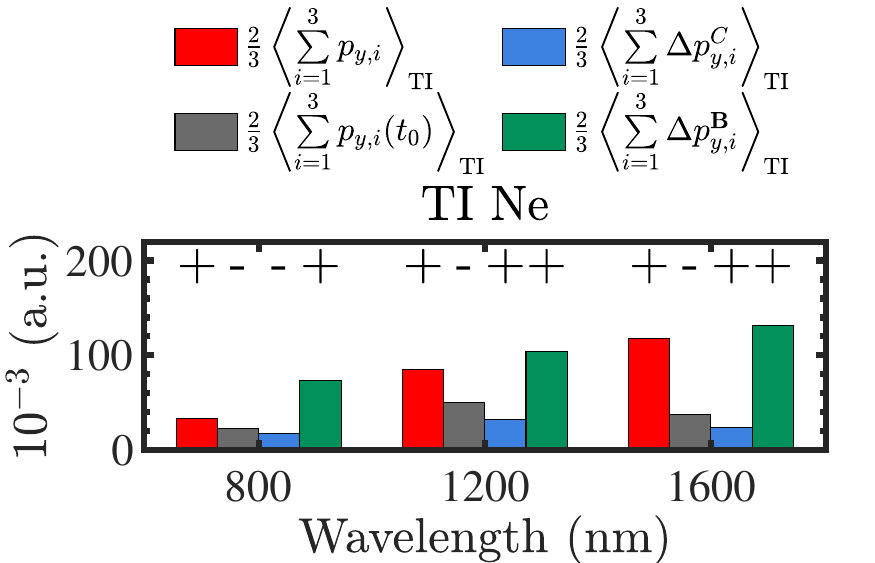}
\caption{For Ne, we show the final average momentum offset per pair of electrons for TI at 800 nm (left bar cluster), 1200 nm (middle bar cluster) and 1600 nm (right bar cluster). We show the momentum offset for TI, $2 / 3\left\langle\sum_{i=1}^3  p_{y, i}\right\rangle$ (red bar), and the contributions due to the initial momentum, $2 / 3\left\langle\sum_{i=1}^3  p_{y, i}(t_0)\right\rangle$ (gray bar), the magnetic field $2 / 3\left\langle\sum_{i=1}^3  \Delta p^\mathbf{B}_{y, i}\right\rangle$ (green bar), and the Coulomb plus effective Coulomb forces, $2 / 3\left\langle\sum_{i=1}^3  \Delta p^C_{y, i}\right\rangle$ (blue bar). The plus (+) or minus (-) sign above each bar denotes a positive or negative value, for the given contribution.}
\label{fig:Offset_all}
\end{figure}

In Fig.~\ref{fig:Offset_all}, for TI of strongly driven Ne, we compute the average sum of the final momenta of all three electrons (red bars) along the direction of light propagation at 1600 nm (right bar cluster), 1200 nm (middle bar cluster)  and  800 nm (left bar cluster).  In what follows, we refer to this  average sum as momentum offset.   All momenta offsets are computed per electron pair, i.e. we multiply the momentum offset by a factor of $2/3$ as follows
\begin{equation}
\begin{aligned}
& \frac{2}{3}\left\langle\sum_{i=1}^3 p_{y, i}\right\rangle_{\mathrm{TI}} \\
& \quad=\left\langle\frac{\left(p_{y, 1}+p_{y, 2}\right)+\left(p_{y, 1}+p_{y, 3}\right)+\left(p_{y, 2}+p_{y, 3}\right)}{3}\right\rangle.
\end{aligned}
\end{equation}
This multiplication allows for a direct comparison of the momentum offset in NSDI of He driven at 800 nm in Ref.~\cite{Emmanouilidou2} and in NSTI of Ne driven at  1600 nm, 1200 nm and 800 nm.
Here, we denote by $p_{y,i}$ the $y$ component (direction of light propagation) of the final momentum of electron $i$. We note that in the dipole approximation  the momentum offset is zero. We find that when Ne is driven by a pulse at 1600 nm and 1200 nm the momentum offsets in TI are $0.115$ a.u. and $0.085$ a.u., respectively, significantly larger than the momentum offset of $0.033$ a.u. at 800 nm.  That is, the momentum offset is enhanced by a factor of 3 when Ne is driven by a mid-infrared laser pulse. Also, the momentum offsets of TI in Ne driven at 1600 nm and  1200 nm are comparable to the large momentum offset of DI in He driven at 800 nm~\cite{Emmanouilidou1,Emmanouilidou2}. 

Next, we investigate   the reason for the enhancement of the momentum offset in TI of Ne when driven by a laser pulse at 1600 nm and 1200 nm. To do so, we  focus on the different contributions to the momentum offset, which for each electron $i$ are given by
 
\begin{equation}
\left\langle p_{y, i}\right\rangle=\left\langle p_{y, i}\left(t_0\right)\right\rangle+\left\langle\Delta p_{y, i}^C\right\rangle+\left\langle\Delta p_{y, i}^{\mathbf{B}}\right\rangle .
\end{equation}
The first term $\left\langle p_{y, i}\left(t_0\right)\right\rangle$ is the average initial momentum, the term $\left\langle\Delta p_{y, i}^C\right\rangle$ is the average momentum change due to the Coulomb forces and  the effective Coulomb forces over the time interval $[t_0,t_f]$;  $t_f$ denotes the end of the time propagation. The last term, $\left\langle\Delta p_{y, i}^{\mathbf{B}}\right\rangle$, denotes the average momentum change due to the magnetic field of the pulse over the  time interval $[t_0,t_f]$. The average sum of these contributions over all electrons  is shown in Fig.~\ref{fig:Offset_all}.

\begin{figure}[t]
	\centering   
\includegraphics[width=\linewidth]{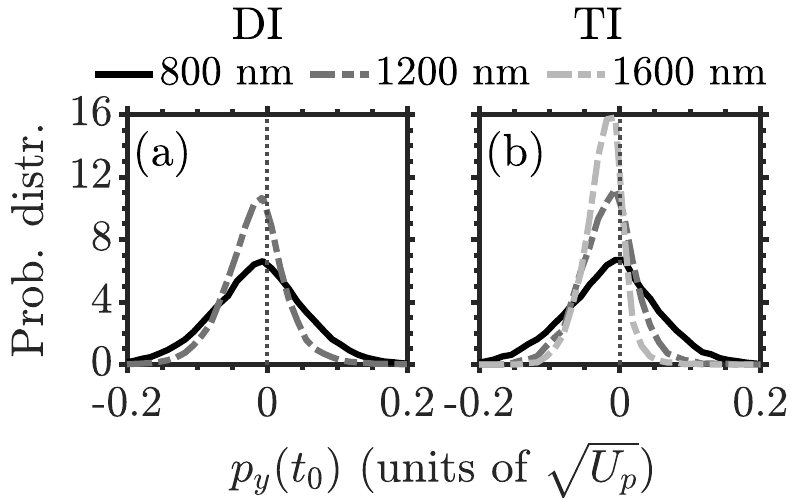}
\caption{For DI (left) and TI (right) of Ne, we plot the  distribution of the $y$ component of the initial momentum (at time $t_0$) of electron 1 at 1600 nm (dash-dotted light gray line for TI), 1200 nm (dash-dotted  gray line) and 800 nm (solid black line). For DI, we only consider the 1200 nm and 800 nm. All distributions are normalized to one.}
\label{fig:position_momentum}
\end{figure}

Fig.~\ref{fig:Offset_all} shows that  the contribution to the momentum offset due to the initial electron momentum, i.e. $2 / 3\left\langle\sum_{i=1}^3  p_{y, i}(t_0)\right\rangle$ (gray bars), is negative at all wavelengths, having a more negative value at 1600 nm and 1200 nm. This shows that nondipole gated ionization is present at all wavelengths, while more significant at 1600 nm and 1200 nm. To further demonstrate the enhancement of nondipole gated ionization at 1600 nm and 1200 nm, in  Fig.~\ref{fig:position_momentum}(b), we plot the distribution of the initial momentum of the recolliding electron.  First, we find that the percentage of TI events with the recolliding electron having negative initial momentum  is $56\%$ at 800 nm, $68\%$ at 1200 nm and  $83\%$ at 1600 nm. Also, we find that  the average sum of the initial momenta of all three electrons  is $-0.024\,\sqrt{U_p}$ at 1600 nm, $-0.018\,\sqrt{U_p}$ at 1200 nm and $-0.011\,\sqrt{U_p}$ at 800 nm.

Also, Fig.~\ref{fig:Offset_all} clearly shows that the dominant contribution to the momentum offset in TI   is due to the magnetic field (green bars) when Ne is driven by a pulse at all three wavelengths. This magnetic-field contribution is roughly two times larger at 1600 nm compared to 800 nm. In the following section, we explain how this enhancement of the magnetic-field contribution to the momentum offset arises with increasing wavelength. 

\begin{figure}[b]
	\centering
\includegraphics[width=\linewidth]{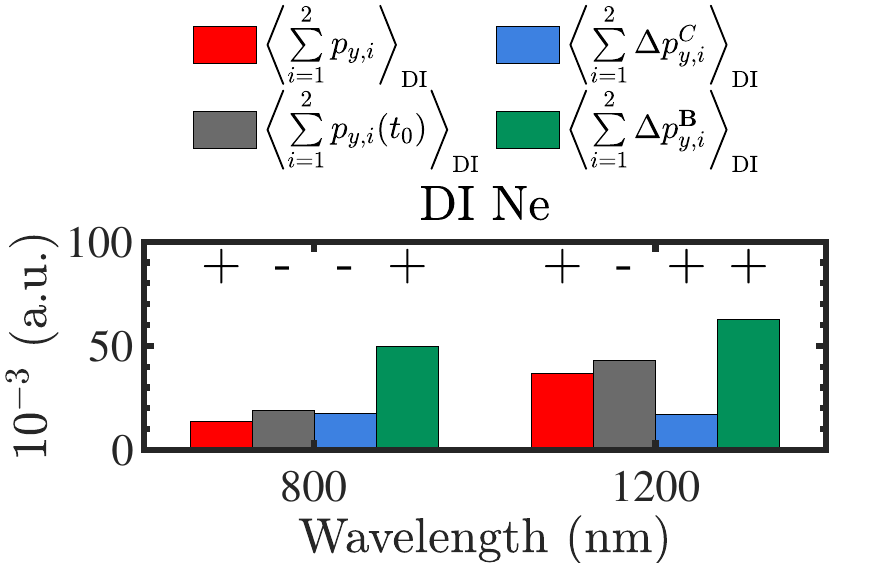}
\caption{For Ne, we show the final average momentum offset  for DI at 800 nm (left bar cluster) and 1200 nm (right bar cluster).  We show the momentum offset for DI, $\left\langle\sum_{i=1}^2  p_{y, i}\right\rangle$ (red bar), and the contributions due to the initial momentum, $\left\langle\sum_{i=1}^2  p_{y, i}(t_0)\right\rangle$ (gray bar), the magnetic field $\left\langle\sum_{i=1}^2 \Delta p^\mathbf{B}_{y, i}\right\rangle$ (green bar), and the Coulomb plus effective forces $\left\langle\sum_{i=1}^2  \Delta p^C_{y, i}\right\rangle$ (blue bar). The plus (+) or minus (-) sign above each bar denotes a positive or negative value, for the given contribution.}
\label{fig:Offset_DI}
\end{figure}

Similar findings hold for the momentum offset in DI of strongly driven Ne. That is, the momentum offset for DI is significantly larger when Ne is driven by a laser pulse at 1200 nm  compared to Ne being driven by a pulse at 800 nm, see  Fig.~\ref{fig:Offset_DI}. We do not consider the 1600 nm for DI of strongly driven Ne, since at 1600 nm we find that only half of the events have a recollision.  This is consistent with a decrease of electron-electron correlation in the recollisions with increasing wavelength, see  section III.B. Moreover, for both wavelengths nondipole gated ionization is present. Indeed, the momentum offset due to the initial electron momentum is negative for both wavelengths in Fig.~\ref{fig:Offset_DI}. Also, the percentage of DI events with the recolliding electron having negative initial momentum is larger compared to positive values for both wavelengths, see Fig.~\ref{fig:position_momentum}(a). 
  Moreover, 
we find that the momentum offsets in DI are smaller compared to the respective ones in TI, compare red bars in Fig.~\ref{fig:Offset_all} and Fig.~\ref{fig:Offset_DI}.  This is consistent with weaker electron-electron correlation in recollisions in DI compared to TI, see also section III.C on pathways of TI and DI. However, as for TI, in DI the momentum offset is mainly due to the contribution of the magnetic field of the laser field. 
\subsection{Contribution to the momentum offset due to the magnetic field}

\subsubsection{Simple model for the contribution to the momentum offset due to the magnetic field}

In this section, we employ a simple model, first developed in Ref.~\cite{katsoulis_nondipole_2023}, to account for the contribution to the momentum offset due to the magnetic field in TI and DI of strongly driven Ne.  We show that this model accounts for  the enhancement of the momentum offset in TI of Ne when driven by a laser pulse at 1600 nm and 1200 nm  compared to when driven by a pulse at 800 nm as well as for the enhancement in DI at 1200 nm. In what follows, we focus  on TI of strongly driven Ne.

For completeness, we briefly describe the simple model developed in Ref.~\cite{katsoulis_nondipole_2023}. The Lorentz force acting upon an electron $i$ in an electromagnetic field is given by 
\begin{equation}
\mathbf{F}_{\mathrm{L}}=-\left[\mathbf{E}\left(y_i, t\right)+\mathbf{p}_i \times \mathbf{B}\left(y_i, t\right)\right].
\end{equation}
The momentum of electron $i$ is given by
\begin{align}\label{eq:momentum2}
\mathbf{p}_{i}(t)&=\mathbf{p}_{i}(t_0)-\int_{t_0}^{t}  \left[ \mathbf{E}(y_i,t') + \mathbf{p}_i \times \mathbf{B}(y_i,t') \right] dt' \nonumber \\
%%%%%%%%%%%%%%%%%%%%%%%%%%%%%%%%%%%%%%%%%%%%
&+ \mathrm{H}(t-t_{\mathrm{rec}})\Delta\mathbf{p}_{i}(t_{\mathrm{rec}}),
\end{align}
with $\mathrm{H}(t-t_{\mathrm{rec}})$ the Heaviside function~\cite{abramowitz1965handbook} and $\Delta\mathbf{p}_{i}(t_{\mathrm{rec}})$ being the momentum change of electron $i$ due to the recollision that occurs upon the return of the recolliding electron to the core.  In this simple model, we do not account for the Coulomb forces between electrons and  between each electron and the core. These forces are accounted for only indirectly  via the term $\Delta\mathbf{p}_{i}(t_{\mathrm{rec}})$.
 
\begin{figure}[t]
	\centering
\includegraphics[width=\linewidth]{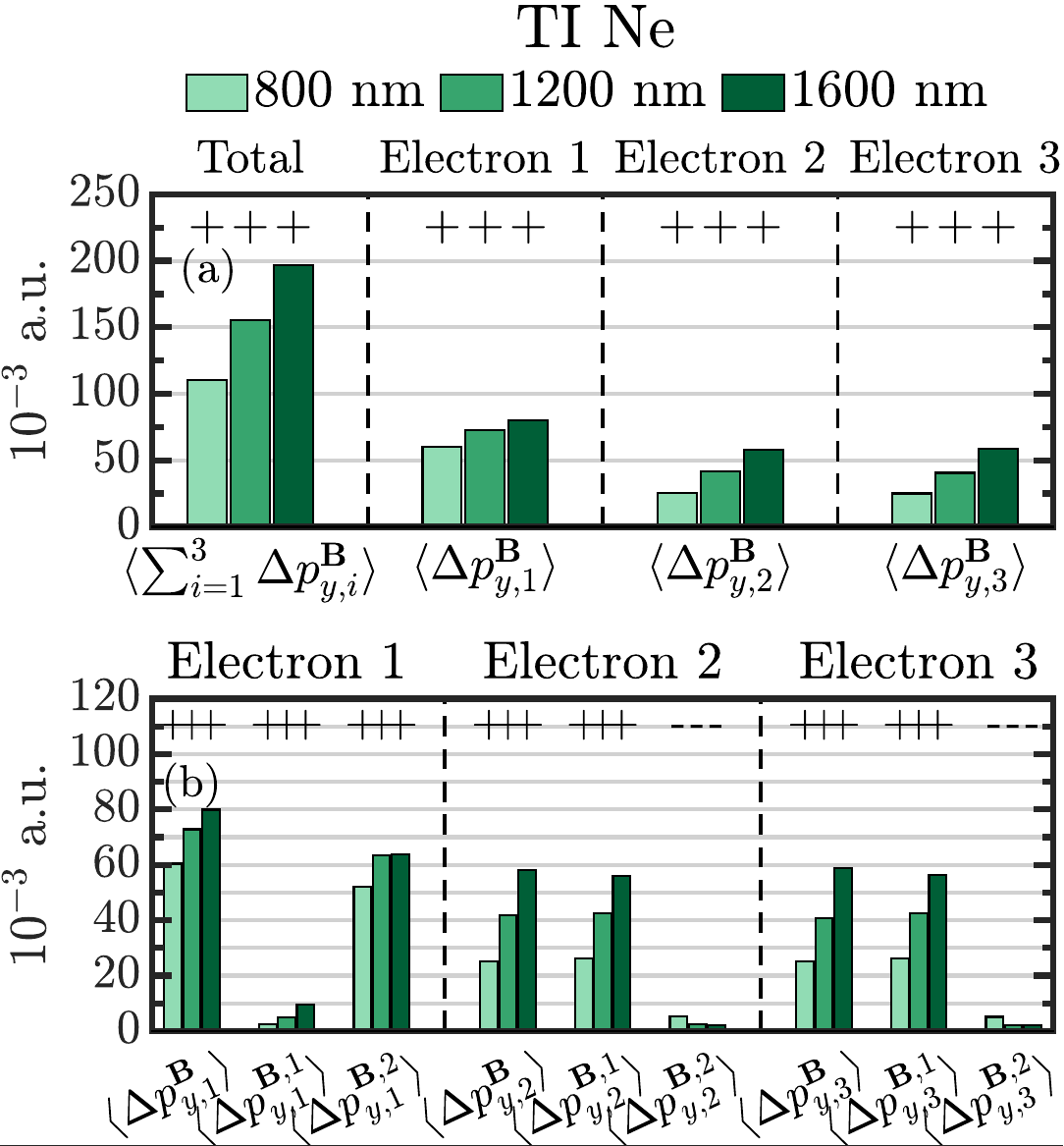}
\caption{For TI of Ne, we show the contribution to the momentum offset due to the magnetic field (a) of all electrons,  $\left\langle \sum_{i=1}^{i=3} \Delta p_{y,i}^{\mathbf{B}} \right\rangle$, as well as of each electron separately, $\left\langle \Delta p_{y,i}^{\mathbf{B}} \right\rangle$, at 1600 nm (dark green), 1200 nm (intermediate green shade) and 800 nm (light green). (b) For each electron, we show the contributions of $\left\langle \Delta p_{y,i}^{\mathbf{B,1}} \right\rangle$ and $\left\langle \Delta p_{y,i}^{\mathbf{B,2}} \right\rangle$ to $\left\langle \Delta p_{y,i}^{\mathbf{B}} \right\rangle$ at 1600 nm, 1200 nm and 800 nm.}
\label{fig:histogram_magnetic}
\end{figure}

\begin{figure}[t]
	\centering
\includegraphics[width=\linewidth]{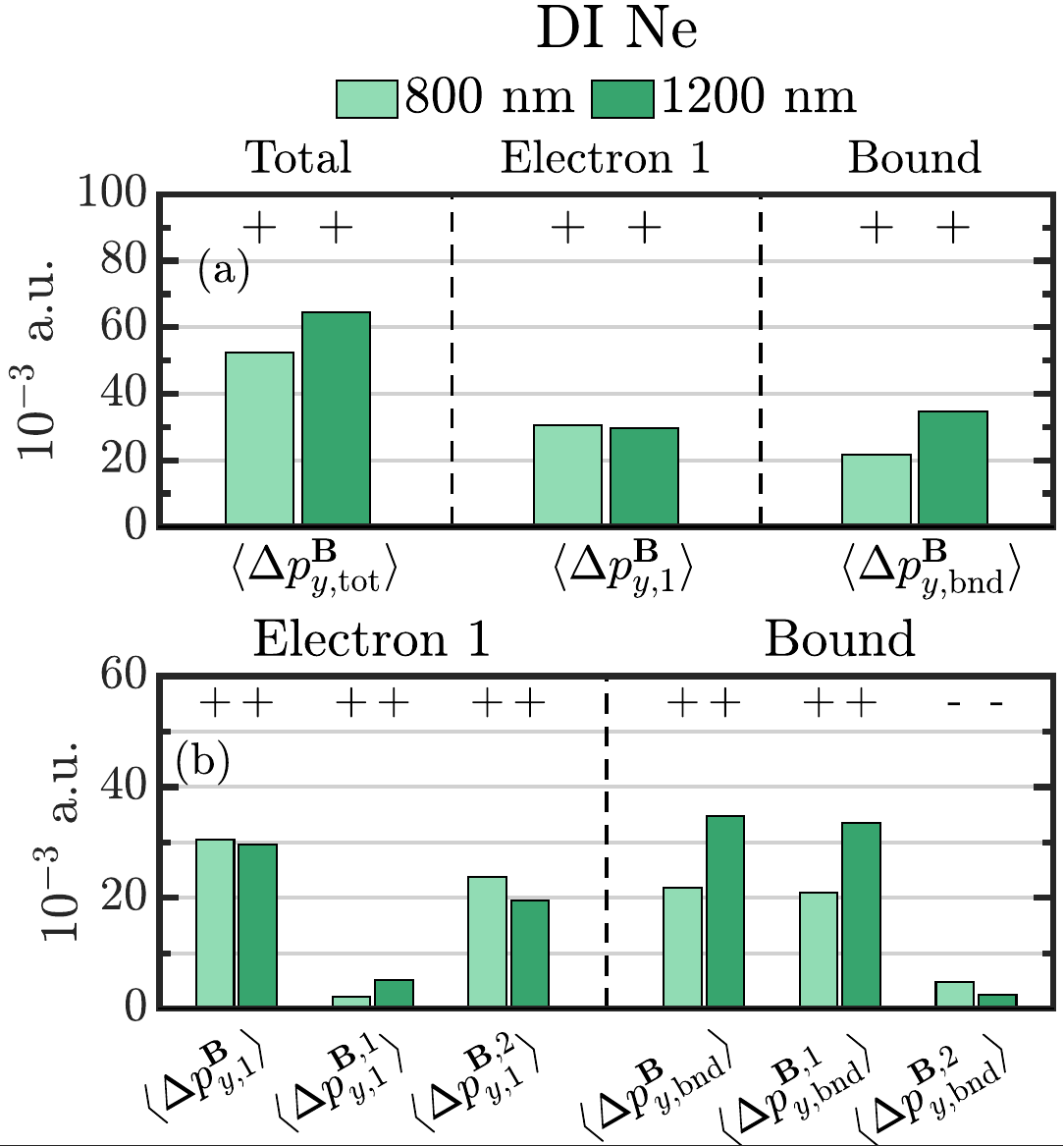}
\caption{For DI of Ne, we show the contribution to the momentum offset due to the magnetic field (a) of the two escaping electrons,  $\left\langle \sum_{i=1}^{i=2} \Delta p_{y,i}^{\mathbf{B}} \right\rangle$, as well as of each of the two escaping electrons separately, $\left\langle \Delta p_{y,i}^{\mathbf{B}} \right\rangle$, at 1200 nm (intermediate green shade) and 800 nm (light green). (b) For each of the two escaping electrons, we show the contributions of $\left\langle \Delta p_{y,i}^{\mathbf{B,1}} \right\rangle$ and $\left\langle \Delta p_{y,i}^{\mathbf{B,2}} \right\rangle$ to $\left\langle \Delta p_{y,i}^{\mathbf{B}} \right\rangle$ at 1200 nm and 800 nm.}
\label{fig:histogram_magneticDI}
\end{figure}

As in the detailed ECBB model, in the simple model, we take the initial momentum of the recolliding electron along the direction of the electric field to be zero, i.e., $p_{z,i}(t_0)=0$.  Also, when using the simple model to compute $\left\langle \Delta p_{y,i}^{\mathbf{B}} \right\rangle$, we neglect terms of order $\mathbf{B}^2$, since the ratio of the magnitudes of the electric and magnetic field is $| E(y_i,t)/B(y_i,t) | = c$, i.e. the magnitude of the electric field is much larger. We also compute the magnetic and electric fields as a function of time at the same position $y_i=0$, i.e., 
\begin{align}
& E\left(y_i, t\right) \approx E(0, t) \equiv E(t), \\
%%%%%%%%%%%%%%%%%%%%%%%%%%%%%%%%%%%%%%%%%%%%
& B\left(y_i, t\right) \approx B(0, t) \equiv B(t) .
\end{align}

Given the above approximations and Eq.~\eqref{eq:momentum2}, we find  that the momentum change due to the magnetic field is given by (see Ref.~\cite{katsoulis_nondipole_2023} for details)
\begin{equation}\label{eq:dpB_a}
\Delta p^{\mathbf{B}}_{y,i}(t_\alpha \rightarrow t) = 
\Delta p^{\mathbf{B},1}_{y,i}(t_{\alpha} \rightarrow t) + 
\Delta p^{\mathbf{B},2}_{y,i}(t_{\mathrm{rec}} \rightarrow t), 
\end{equation}
with
\begin{subequations}
\begin{align}
\Delta p^{\mathbf{B},1}_{y,i}(t_{\alpha} \rightarrow t)
&= \int_{t_{\alpha}}^{t} \! \left[ \! \int_{t_{\alpha}}^{t'} E(t'')\,dt'' \right]\, B(t')\,dt', \label{eq:dpB_b} \\[6pt]
%%%%%%%%%%%%%%%%%%%%%%%%%%%%%%%%%%%%%%%%%%%%
\Delta p^{\mathbf{B},2}_{y,i}(t_{\mathrm{rec}} \rightarrow t)
&= - \Delta p_{z,i}(t_{\mathrm{rec}}) \int_{t_{\mathrm{rec}}}^{t} B(t')\,dt', \label{eq:dpB_c}
\end{align}
\end{subequations}
where $t_{\alpha}=t_{0}$ for the recolliding electron, and $t_{\alpha}=t_{ \mathrm{ion} }$ for each of the  initially bound electrons. We note that in Ref.~\cite{katsoulis_nondipole_2023}, $t_{\alpha}=t_{ \mathrm{rec} }$ for the initially bound electrons. However, in the current work, we find the ionization time to be more relevant compared to the recollision time for the  momentum change due to the magnetic field, $\Delta p_{y, i}^{\mathbf{B}}$, for the initially bound electrons. The reason is that the recollisions are softer at 1600 nm and 1200 nm, compared to 800 nm, as we show in what follows.

 %As was the case in Ref.~\cite{katsoulis_nondipole_2023} we find that the term $\Delta p_{y,i}^{\mathbf{B},2}(t_{\mathrm{rec}} \to t_f)$ contributes the most to $\Delta p_{y,i}^{\mathbf{B}}(t_{0} \to t_f)$ for a recolliding electron; for an initially bound electron the term $\Delta p_{y,i}^{\mathbf{B},1}(t_{\mathrm{ion}} \to t_f)$ contributes the most. Moreover, both these terms were shown to be positive for the recolliding and bound electrons respectively, see Ref.~\cite{katsoulis_nondipole_2023}. 

In Fig.~\ref{fig:histogram_magnetic}(a),  we show the contributions to the momentum offset due to the magnetic field  of all electrons (Total) as well as of each electron $i$ separately. 
 Fig.~\ref{fig:histogram_magnetic}(a) clearly shows that the contribution to the momentum offset due to the magnetic field, $\left\langle \sum_{i=1}^{i=3} \Delta p_{y,i}^{\mathbf{B}} \right\rangle$,  is enhanced at 1200 nm and at 1600 nm mainly due to the bound electrons. Indeed, 
 $\left\langle \Delta p_{y,1}^{\mathbf{B}} \right\rangle$ for the recolliding electron (electron 1)  increases slightly with increasing wavelength. However,  $\left\langle \Delta p_{y,2/3}^{\mathbf{B}} \right\rangle$ for the bound electrons increases significantly with increasing wavelength and  is at least twice as large at 1600 nm compared to 800 nm. %We find that at 1600 nm the bound electrons account  for 59\% of $\left\langle \sum_{i=1}^{i=3} \Delta p_{y,i}^{\mathbf{B}} \right\rangle$, while at 800 nm the bound electrons account only for 45 \% of  $\left\langle \sum_{i=1}^{i=3} \Delta p_{y,i}^{\mathbf{B}} \right\rangle$.
 
Next, for each electron, we consider  the contributions of the terms $\left\langle \Delta p_{y,i}^{\mathbf{B},1} \right\rangle$ and $\left\langle \Delta p_{y,i}^{\mathbf{B},2} \right\rangle$, defined in Eq.~\eqref{eq:dpB_b} and  Eq.~\eqref{eq:dpB_c}, respectively, to $\Delta p_{y, i}^{\mathbf{B}}$. This will allow us to understand  the reason that for 1600 nm and 1200 nm 
 $\left\langle \Delta p_{y,i}^{\mathbf{B}} \right\rangle$ significantly increases compared to 800 nm  for the bound electrons (electrons 2 and 3),  while it barely increases  for the recolliding electron (electron 1). Fig.~\ref{fig:histogram_magnetic}(b) clearly shows that the term $\left\langle \Delta p_{y,i}^{\mathbf{B},1} \right\rangle$ is the dominant contribution to $\left\langle \Delta p_{y,i}^{\mathbf{B}} \right\rangle$ for each bound electron, while the term $\left\langle \Delta p_{y,i}^{\mathbf{B},2} \right\rangle$ almost solely accounts for  $\left\langle \Delta p_{y,i}^{\mathbf{B}} \right\rangle$ for the recolliding electron. Hence, we next study how the terms $\left\langle \Delta p_{y,i}^{\mathbf{B},1} \right\rangle$ for the bound electrons and $\left\langle \Delta p_{y,i}^{\mathbf{B},2} \right\rangle$  for the recolliding electron vary with wavelength. 
 
 For DI of strongly driven Ne, similar findings hold as for TI  for the contributions to the momentum offset due to the magnetic field  of all electrons (Total) as well as of each electron $i$ separately, see Fig.~\ref{fig:histogram_magneticDI}.
 
\subsubsection{Contribution to the momentum offset due to the magnetic field for the  bound electrons}

 \begin{figure}[t]
	\centering
\includegraphics[width=\linewidth]{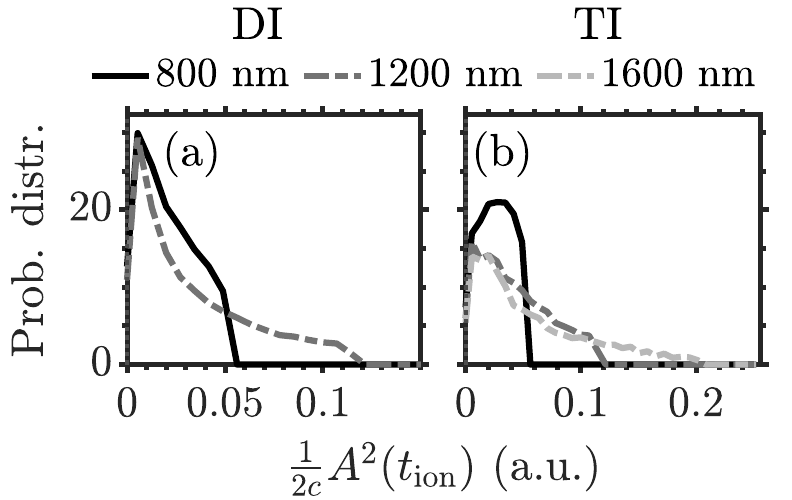}
\caption{For DI (left) and TI (right) of strongly driven Ne, we plot  the probability distribution of  $ \frac{1}{2c} A^2(t_{\mathrm{ion}})$ for the bound electrons at 1600 nm (dash-dotted light gray line for TI), 1200 nm (dash-dotted gray line) and  800 nm (solid black line). All distributions are normalized to one.}
\label{fig:vector_potential_sq}
\end{figure}

Assuming that the momentum of each bound electron just before recollision is zero, i.e., $p_i(t_{\alpha}) \approx 0$, we have shown in Ref.~\cite{katsoulis_nondipole_2023}  that 
\begin{equation}\label{eq:momentumgainsBound_rec}
\Delta p_{y,i}^{\mathbf{B},1}(t_{\mathrm{rec}} \to t_f) = \frac{1}{2c} A^2(t_{\mathrm{rec}}).
\end{equation}
Here, the ionization time is more relevant for the bound electrons, as previously mentioned. Hence, we substitute  $t_{\mathrm{rec}}$ with $t_{\mathrm{ion}}$ in Eq.~\eqref{eq:momentumgainsBound_rec}.
 In Fig.~\ref{fig:vector_potential_sq}(b), for TI of strongly driven Ne, we plot the distribution of $\frac{1}{2c}A^2(t_{\mathrm{ion}})$ for the bound electrons at 1600 nm, 1200 nm and  800 nm. This distribution shifts to larger values with increasing wavelength.  
 We also find that the average values of the distribution  is 0.026 a.u. at 800 nm, 0.043 a.u. at 1200 nm and  0.056 a.u. at 1600 nm.  These  values are consistent with the values of $\Delta p_{y,i}^{\mathbf{B}}$ for the bound electrons in Fig.~\ref{fig:histogram_magnetic}(a), reinforcing the validity of our simple model.
 
 Indeed, it is  expected that  the term $ \frac{1}{2c} A^2(t_{\mathrm{ion}})$ increases with increasing wavelength, since the amplitude of the vector potential is given by $E_0/\omega$. For instance, one would expect that  the term $ \frac{1}{2c} A^2(t_{\mathrm{ion}})$ is four times larger at 1600 nm compared to 800 nm. However, this  assumes that the  bound electrons ionize at the same time at both wavelengths. We find that the distributions of the ionization times differ as a function of wavelength (not shown). Namely,  ionization times of the bound electrons shift from being closer to a zero of the electric field at 800 nm to being closer to extrema of the electric field at  1600 nm. This is consistent with  recollisions becoming softer with increasing wavelength, see next section.  As a result, at 1600 nm, the values of the vector potential $A(t_{\mathrm{ion}})$ are smaller than its maximum value of  $E_0/\omega$. This explains the increase of the average value of $ \frac{1}{2c} A^2(t_{\mathrm{ion}})$ by a factor of two instead of four at 1600 nm compared to 800 nm.
 
It is important to note that the significant  momentum offset of the bound electrons is due to the electron-electron correlation in recollisions. This is best demonstrated in DI of Ne when driven by a pulse at 1600 nm,  since for these field parameters half of the DI events have no recollisions. We find that the momentum offset of the bound electron is significantly larger for the events with recollisions compared to the ones without recollisions.

\subsubsection{Contribution to the momentum offset due to the magnetic field for the  recolliding electron}

\begin{figure}[t]
	\centering
\includegraphics[width=\linewidth]{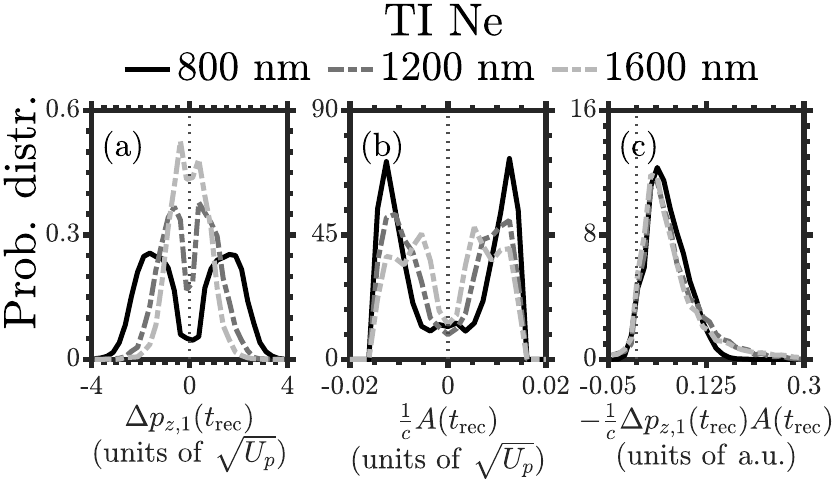}
\caption{For TI of strongly driven  Ne, for the recolliding electron,  we plot the probability distributions of (a)  $\Delta p_{z,i}(t_{\mathrm{rec}})$, (b)  $A(t_{\mathrm{rec}})/c$ and (c) $-\Delta p_{z,i}(t_{\mathrm{rec}}) A(t_{\mathrm{rec}})/c$  at 1600 nm (dash-dotted light gray line), 1200 nm (dash-dotted gray line) and 
800 nm (solid black line).  $\Delta p_{z,i}(t_{\mathrm{rec}})$ is obtained using the momenta of the recolliding electron at times $\mathrm{t_{rec}+T/50}$ and $\mathrm{t_{rec}-T/50}$. All distributions are normalized to one.}
\label{fig:pz_change_magnetic_integral}
\end{figure}

\begin{figure}
	\centering
\includegraphics[width=\linewidth]{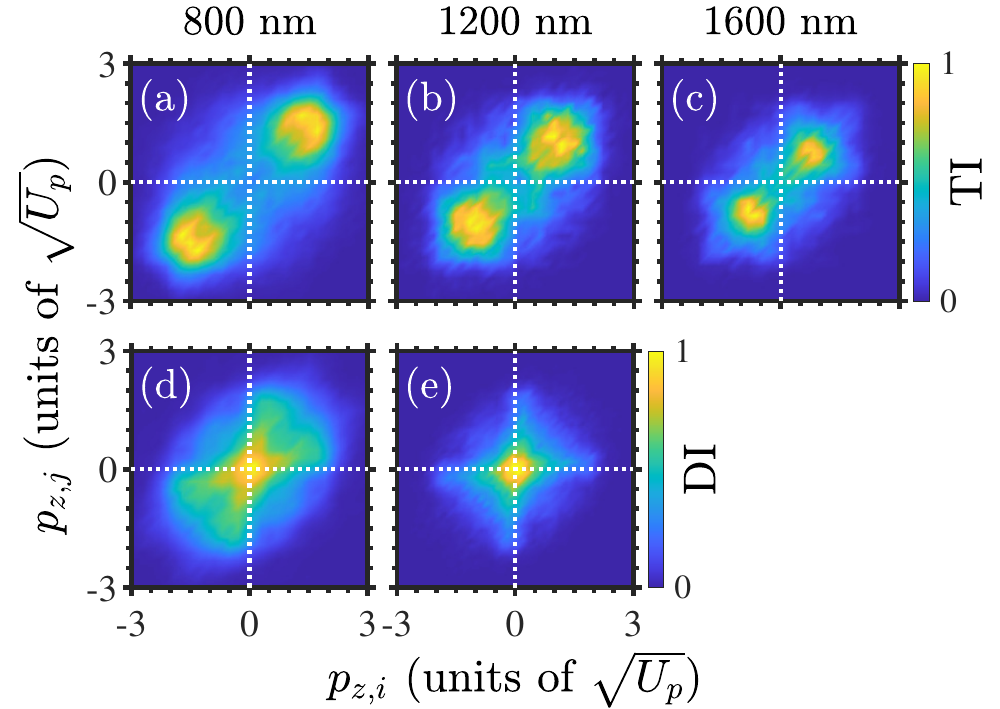}
\caption{For TI (top row) and  DI (bottom row) of strongly driven Ne, symmetrized correlated momenta $p_z$ for all pairs of escaping electrons at  800 nm (left), 1200 nm (middle) and at 1600 nm (right for TI). }
\label{fig:Correlated_momenta}
\end{figure}

For the recolliding electron, see details in  Ref.~\cite{katsoulis_nondipole_2023}, we find that
\begin{equation}\label{eq:Appendix_MG_inc_v2}
\Delta p_{y,i}^{\mathbf{B},2}(t_{\mathrm{rec}} \to t_f) = -\frac{1}{c} \Delta p_{z,i}(t_{\mathrm{rec}}) A(t_{\mathrm{rec}}).
\end{equation}
Next,  for the recolliding electron, we  explain  the reason  $\Delta p_{y,i}^{\mathbf{B},2}(t_{\mathrm{rec}} \to t_f)$   does not significantly change as a function of the wavelength, see Fig.~\ref{fig:histogram_magnetic}(b). To do so,  we study how each of the terms  $\Delta p_{z,i}(t_{\mathrm{rec}})$ and $A(t_{\mathrm{rec}})/c$ changes with wavelength for TI of strongly driven Ne. In Fig.~\ref{fig:pz_change_magnetic_integral}(b) we  find that the distribution of $A(t_{\mathrm{rec}})/c$ in units of $\sqrt{U_{p}}$ extends to increasingly larger values as the wavelength decreases from 1600 nm to 800 nm. This is consistent with recollisions occurring  closer to a zero of the electric field at 800 nm compared to 1600 nm, resulting in larger values of the vector potential. Also, Fig.~\ref{fig:pz_change_magnetic_integral}(a) shows that the distribution of $\Delta p_{z,i}(t_{\mathrm{rec}})$  in units of $\sqrt{U_{p}}$ decreases with increasing wavelength. That is,  the recolliding electron transfers less energy to the bound electrons during recollision with increasing wavelength. 
This larger transfer of momentum along the polarization direction during recollisions and the resulting stronger electron-electron correlation is what we refer to as strong recollisions
  at 800 nm. The smaller transfer of momentum during recollisions and the resulting weaker electron-electron correlation is what we refer to as softer recollisions at 1200 nm and even softer at 1600 nm. Fig.~\ref{fig:Correlated_momenta} further illustrates the softer  recollisions for 1200 nm and 1600 nm  for TI and DI, since the magnitude of the momenta of the electrons are decreasing with increasing wavelength.

However, for increasing wavelength, the decrease of the average and peak values of both distributions of  $\Delta p_{z,i}(t_{\mathrm{rec}})$  and $A(t_{\mathrm{rec}})/c$ in units of $\sqrt{U_{p}}$ is compensated by the increase of  $\sqrt{U_{p}}$. Indeed, Fig.~\ref{fig:pz_change_magnetic_integral}(c) shows that the distribution of $-\frac{1}{c} \Delta p_{z,i}(t_{\mathrm{rec}}) A(t_{\mathrm{rec}})$ is similar for all three wavelengths with average values of 0.055 a.u. at 800 nm,  0.063 a.u. at 1200 nm and 0.062 a.u. at 1600 nm. These values are close to the ones in  Fig.~\ref{fig:histogram_magnetic}(a) for electron 1, which reinforces the validity of our simple model.

%In contrast, the term $\Delta p_{z,i}(t_{\mathrm{rec}})$ for the recolliding electron decreases at 1600 nm compared to 800 nm. 
%This is clearly shown in Fig.~\hyperref[fig:pz_change_magnetic_integral]{\ref*{fig:pz_change_magnetic_integral}(a)}. Indeed, at 1600 nm, the momentum change of the recolliding electron along the polarization direction of the electric field is significantly smaller compared to 800 nm. This means that the recollisions are softer at 1600 nm since there is less energy and hence momentum transfer from the recolliding electron to the bound electrons. These softer recollisions are also consistent with the ionization times for the bound electrons being closer to extrema rather than zeros of the electric field at 1600 nm compared  to 800 nm, as discussed above. We further illustrate that the recollisions are softer at 1600 nm compared to 800 nm by plotting the correlated momenta along the polarization axis of the electric field in Fig.~\ref{fig:Correlated_momenta}. Comparing  Figs.~\ref{fig:Correlated_momenta} (a) and (b), we find that the magnitude of the momenta of the electrons are significantly larger at 800 nm compared to 1600 nm. We find that the average value of  the distribution of $-\frac{1}{c} \Delta p_{z,i}(t_{\mathrm{rec}}) A(t_{\mathrm{rec}})$, shown in  Fig.~\ref{fig:pz_change_magnetic_integralt}, is 0.064 a.u. at 1600 nm and 0.056 a.u. at 800 nm. These average values are close to the values of $\Delta p_{y,i}^{\mathbf{B}}$ for the recolliding electron in Fig.~\hyperref[fig:histogram_magnetic]{\ref*{fig:histogram_magnetic}(a)}, re-enforcing the validity of our simple model.

\begin{figure}[b]
	\centering
\includegraphics[width=\linewidth]{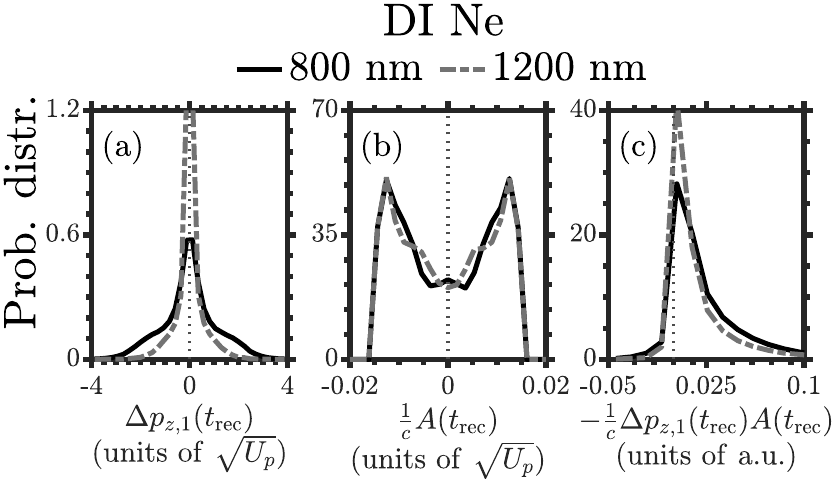}
\caption{For DI of strongly driven  Ne, for the recolliding electron,  we plot the probability distributions of (a)  $\Delta p_{z,i}(t_{\mathrm{rec}})$, (b)  $A(t_{\mathrm{rec}})/c$ and (c) $-\Delta p_{z,i}(t_{\mathrm{rec}}) A(t_{\mathrm{rec}})/c$   at 1200 nm (dash-dotted gray line) and 
800 nm (solid black line).  $\Delta p_{z,i}(t_{\mathrm{rec}})$ is obtained using the momenta of the recolliding electron at times $\mathrm{t_{rec}+T/50}$ and $\mathrm{t_{rec}-T/50}$. All distributions are normalized to one.}
\label{fig:pz_change_magnetic_integralDI}
\end{figure}

Similar findings hold for the momentum offset of the recolliding electron in DI of strongly driven Ne, see  Fig.~\hyperref[fig:pz_change_magnetic_integralDI]{\ref*{fig:pz_change_magnetic_integralDI}}. 

\begin{figure}[t]
	\centering
\includegraphics[width=\linewidth]{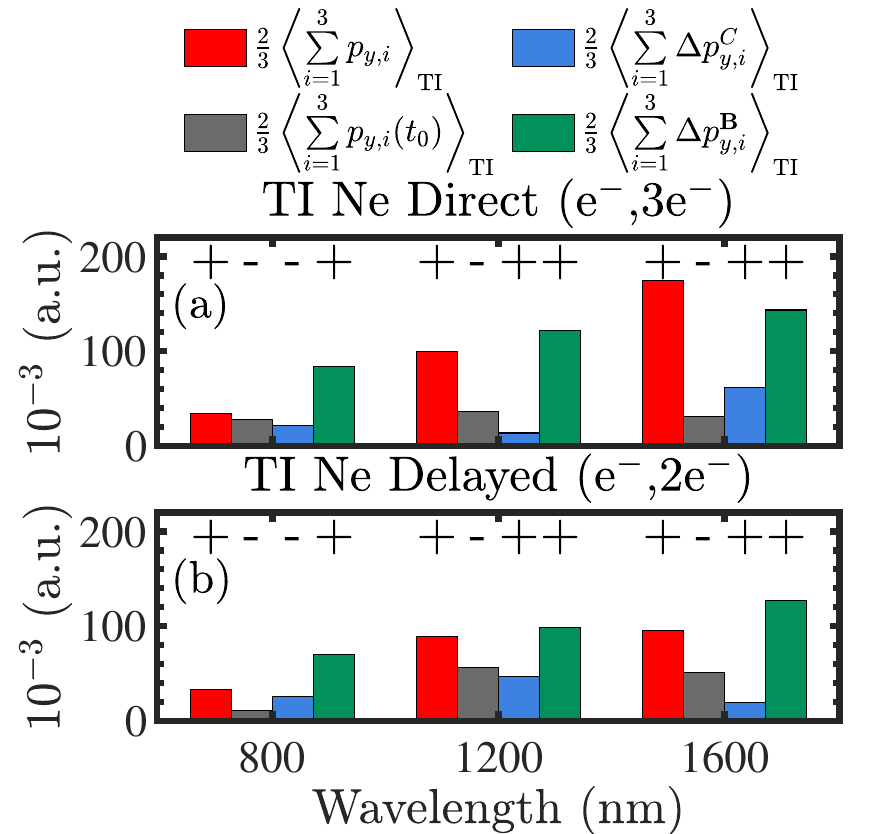}
\caption{For TI of  Ne, we show the final average momentum offset per pair of electrons at 800 nm (left bar cluster), 1200 nm (middle bar cluster) and 1600 nm (right bar cluster). We show the momentum offset $2 / 3\left\langle\sum_{i=1}^3  p_{y, i}\right\rangle$ (red bar) and the contributions due   to the initial momentum $2 / 3\left\langle\sum_{i=1}^3  p_{y, i}(t_0)\right\rangle$ (gray bar), the magnetic field $2 / 3\left\langle\sum_{i=1}^3 \Delta  p^\mathbf{B}_{y, i}\right\rangle$ (green bar), and the Coulomb and effective Coulomb forces $2 / 3\left\langle\sum_{i=1}^3 \Delta  p^C_{y, i}\right\rangle$ (blue bar). The plus (+) or minus (-) sign above each bar denotes a positive or negative value,  for the given contribution. We show  the momentum offsets for the (a) direct and (b)  delayed pathways.}
\label{fig:Direct_delayed_offsetTI}
\end{figure}

\subsection{Momentum offset of the TI pathways as a function of pulse wavelength}\label{Section:Positive_mom_offset_pathway}

\begin{figure}[t]
	\centering
\includegraphics[width=\linewidth]{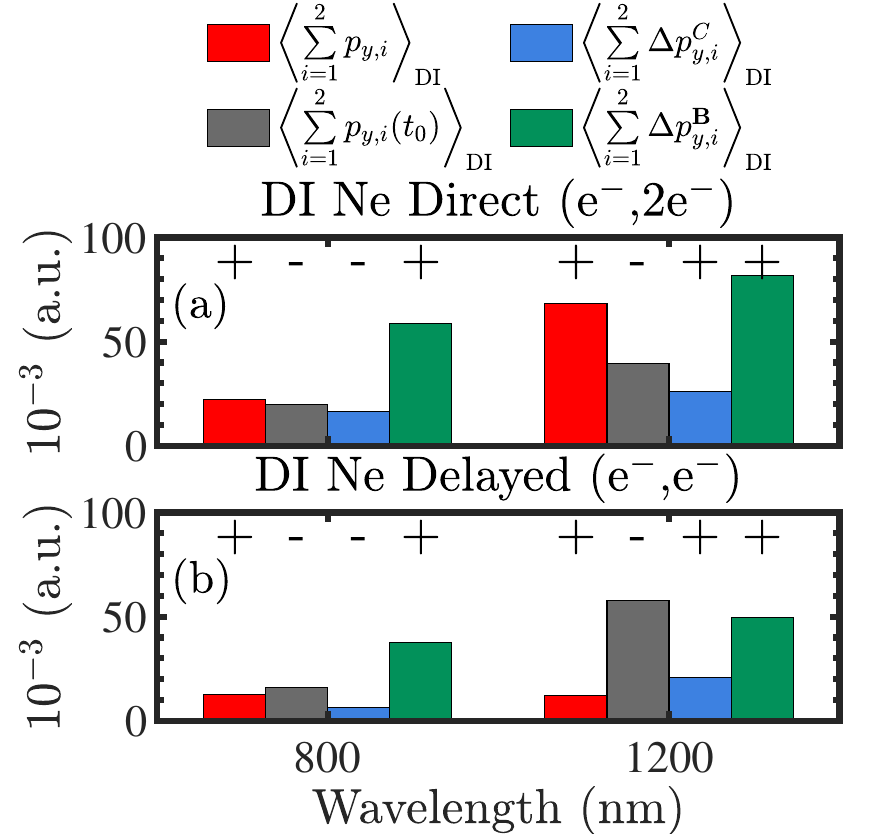}
\caption{For DI of Ne, we show the final average momentum offset  at 800 nm (left bar cluster) and 1200 nm (right bar cluster). We show the momentum offset $\left\langle\sum_{i=1}^2  p_{y, i}\right\rangle$ (red bar) and the contributions due   to the initial momentum $\left\langle\sum_{i=1}^2  p_{y, i}(t_0)\right\rangle$ (gray bar), the magnetic field $\left\langle\sum_{i=1}^2 \Delta  p^\mathbf{B}_{y, i}\right\rangle$ (green bar), and the Coulomb and effective Coulomb forces $\left\langle\sum_{i=1}^2 \Delta  p^C_{y, i}\right\rangle$ (blue bar). The plus (+) or minus (-) sign above each bar denotes a positive or negative value,  for the given contribution. We show  the momentum offsets for the (a) direct and (b)  delayed pathways.}
\label{fig:Direct_delayed_offset}
\end{figure}

Next, we address the dependence of  the momentum offset of the TI pathways  on the  pulse wavelength.  For the TI pathways, we employ the notation $(e^-, ne^-)$, where $n$ denotes the number of electrons that escape shortly after recollision. In the direct $(e^-,3e^-)$ pathway, the recolliding electron transfers sufficient energy to liberate all three electrons, which escape soon after recollision. In the delayed $(e^-,2e^-)$ pathway of TI, only two electrons escape shortly after recollision, while the other electron escapes  at a later time due to the laser field. Hence, in the delayed $(e^-,2e^-)$ pathway recollisions are weaker compared to the direct $(e^-,3e^-)$ pathway of TI. That is, the transfer of energy from the recolliding electron to the bound electrons,  $\Delta p_{z,i}(t_{\mathrm{rec}})$, is larger for the direct pathway of TI. For the recolliding electron the contribution to the momentum offset due to the magnetic field depends on $\Delta p_{z,i}(t_{\mathrm{rec}})$, see  Eq. (\ref{eq:dpB_c}). As a result,  we expect that the momentum offset is larger for the direct rather than the delayed pathway of TI.
 We find that this is indeed the case at all three wavelengths, see
 Fig.~\ref{fig:Direct_delayed_offsetTI}. 
 
In addition, we find that the momentum offset increases for both the direct and delayed pathways of TI with increasing wavelength. Indeed, for each TI pathway the dominant contribution to the momentum offset is due  to the magnetic field (Fig.~\ref{fig:Direct_delayed_offsetTI}) as is the case for all TI events (Fig.~\ref{fig:Offset_all}). Hence,  as discussed in previous sections, the momentum offset due to the magnetic field of the bound electrons is the main reason that the momentum offset increases with wavelength for the TI pathways.
Similar results are obtained for the pathways of DI of strongly driven Ne, see  Fig.~\ref{fig:Direct_delayed_offset}.

Furthermore, we comment on the sign change with wavelength  of the contribution to the momentum offset of the Coulomb and effective Coulomb forces (blue bar) for TI (Fig.~\ref{fig:Offset_all}), for DI (Fig.~\ref{fig:Offset_DI}), for the pathways of TI (Fig.~\ref{fig:Direct_delayed_offsetTI}) and for the pathways of DI (Fig.~\ref{fig:Direct_delayed_offset}). 
 Namely, we find that this contribution has a negative sign at 800 nm while it is positive at 1200 nm and 1600 nm. This is consistent with weaker recollisions for the two larger wavelengths, resembling the glancing recollisions of He driven by a laser pulse at 800 nm where the contribution of the Coulomb forces was also found to be positive  \cite{Emmanouilidou1,Emmanouilidou2}.

 \begin{figure}[t]
	\centering
\includegraphics[width=\linewidth]{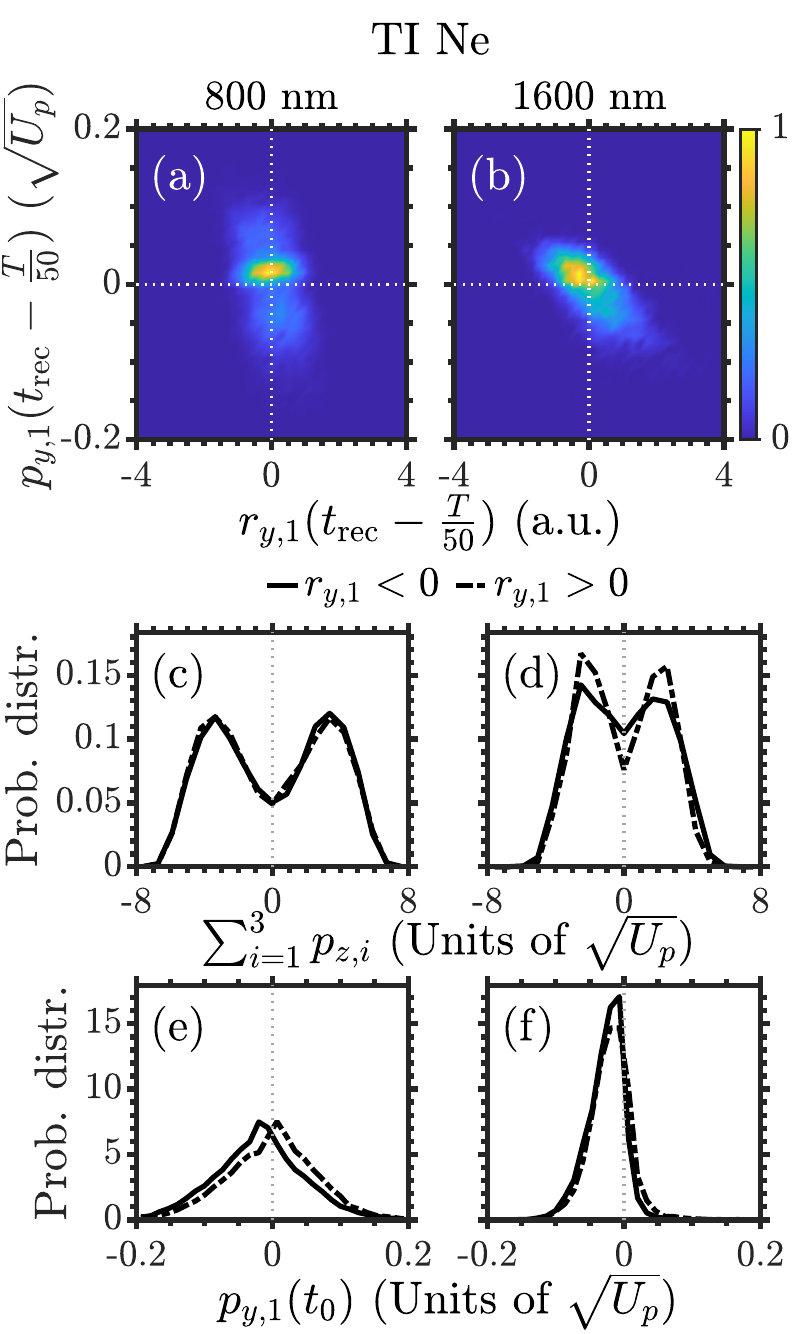}
\caption{For TI of strongly driven Ne, first row: the probability distribution of TI events as a function of the $y$ component of the momentum, $p_{y,1}$ and position, $r_{y,1}$, of electron 1 at time $t_{\mathrm{rec}}-T/50$ at (a) 800 nm and (b) 1600 nm. The momentum is in units of $\sqrt{U_p}$ and the position in a.u. Each distribution is normalized to its maximum value; second row: the sum of the three  final electron momenta p$_{z}$ for TI events with $r_{y,1}<0$ (solid black line) and with $r_{y,1}>0$ (dashed black line); third row: the initial momentum of the recolliding electron along the y-axis for TI events with $r_{y,1}<0$ (solid black line) and with $r_{y,1}>0$ (dashed black line).  The distributions in the second and third row are normalized to one.}
\label{fig:py_ry_before_after}
\end{figure}

Next, using Fig.~\ref{fig:py_ry_before_after}, we explain how the contribution of the Coulomb and effective Coulomb forces to the momentum shift transitions from negative at 800 nm to positive at larger wavelengths, such as 1600 nm considered in  Fig.~\ref{fig:py_ry_before_after}. In this latter figure, we plot in the first row  the probability distribution of TI events as a function of the momentum p$_{y,1}$ and the position r$_{y,1}$ of  electron 1 just before recollision at time $t_{\mathrm{rec}}-T/50$. For TI events with $r_{y,1}<0$  the Coulomb potential between electron 1 and the core contributes with a plus sign to the momentum of electron 1, while it is the other way around for events with  $r_{y,1}>0$. Comparing 800 nm (a) with 1600 nm (b), we find that for TI events with $r_{y,1}<0$  the momentum of electron 1, $p_{y,1}$, overall reduces in magnitude at 1600 nm while this is not  the case for TI events with  $r_{y,1}>0$. Hence,  at 1600 nm, electron 1 spends longer time close to the core when $_{y,1}<0$,  resulting  in a positive contribution of the Coulomb forces to the momentum shift. The reason that at 1600 nm  $p_{y,1}$ reduces overall for TI events with $r_{y,1}<0$ but not for events with $r_{y,1}>0$ is that for the former events the electron-electron correlation is weaker. Indeed,  this is shown in the  second row of Fig.~\ref{fig:py_ry_before_after} by plotting the distribution of the sum of all final electron momenta $p_{z}$. At 800 nm both distributions for $r_{y,1}<0$ and $r_{y,1}>0$  in Fig.~\ref{fig:py_ry_before_after} (c) are the same, suggesting equal strength of electron-electron correlation. However, at 1600 nm, Fig.~\ref{fig:py_ry_before_after} (d)  shows that TI events with $r_{y,1}<0$ are less correlated, i.e. more events have  sum of the momenta  around zero. Since at 1600 nm  the TI events are less correlated for $r_{y,1}<0$,  this implies that the momentum of electron 1, $p_{y,1}$,  does not change significantly due to the recollision. This results in smaller overall momenta $p_{y,1}$ for $r_{y,1}<0$.  Indeed, in Fig.~\ref{fig:py_ry_before_after}  (b) the  momenta in the 2nd and 3rd quadrants are smaller overall than  the momenta in the 1st and 4th quadrants. The above are consistent with the contribution of the Coulomb forces to the momentum shift being positive for larger wavelengths. Note that we find similar results for DI and the separate pathways of TI and DI.

Finally, we explain why the contribution of the Coulomb forces to the momentum shift is negative at 800 nm. The reason is that TI events with  $r_{y,1}>0$ have, in addition to other possible values, momenta $p_{y,1}$ with small positive magnitude, see 1st quadrant in  Fig.~\ref{fig:py_ry_before_after} (a). %TI events with  $\mathrm{r_{y,1}<0}$ can have a combination of  positive or negative small magnitude just before recollision.
 Indeed, events with both negative and positive $r_{y,1}$ experience equally strong electron-electron correlation (see Fig.~\ref{fig:py_ry_before_after} (c)). Hence, at 800 nm, for electron 1 to end up just before recollision with positive $r_{y,1}$ is for $p_{y,1}$ to be  positive just before recollision. This is also supported by the distributions in the third row of Fig.~\ref{fig:py_ry_before_after}, which show that the initial momentum of electron 1 is mostly positive for TI events with $r_{y,1}>0$ just before recollision, particularly so at 800 nm. 
 For  $r_{y,1}>0$, the small positive $p_{y,1}$ is opposite in sign to the negative contribution of the Coulomb force  between electron 1 and the nucleus, resulting in electron 1 spending more time around the core for $r_{y,1}>0$. Hence, the contribution of the Coulomb forces to the momentum shift at 800 nm is negative. We obtain similar results for DI and for the separate pathways of TI and DI.

\section{\label{sec:Conclusions}Conclusions}

In conclusion,  we employed the recently extended ECBB three-dimensional semiclassical model to investigate how nondipole effects change with increasing wavelength in TI and DI of strongly driven Ne. For TI, we have found that the average sum of the final momenta of all three escaping electrons along the direction of light propagation, which we refer to as momentum offset, significantly increases with increasing wavelength. We have found that the momentum offset increases  for all TI events as well as for the direct and delayed pathways of TI. Moreover, using a simple model, we attribute this significant increase of the momentum offset when Ne is driven by laser pulses at 1200 nm and 1600 nm compared to when driven at 800 nm  to the contribution of the effect of the magnetic field on  the bound electrons. This finding concerning the momentum offset as well as   the fact that the TI probability  decreases only by half at 1200 nm compared to 800 nm suggests that 1200 nm is an appropriate wavelength for experimentally measuring the momentum offset related to correlated three-electron escape.

\begin{acknowledgments}
The authors acknowledge the use of the UCL Myriad High Performance Computing Facility (Myriad@UCL), the use of the UCL Kathleen High Performance Computing Facility (Kathleen@UCL), and associated support services in the completion of this work.
\end{acknowledgments}

\hspace{2cm}
\appendix
\section{\label{Appendix_A} Identifying the recollision times }

We obtain the TI and DI events using a code that implements the ECBB model described briefly in Sec.~\ref{sec:Method} and in detail in Refs.~\cite{Agapi3electron,praill_three-electron_2026}. Once these events have been obtained, we perform a detailed analysis using a separate set of codes, where recollisions are identified as follows.

We monitor the interelectronic potential energy for all electron pairs $(i,j)$ during the time intervals in which one electron in the pair is quasifree and the other one is bound. The local maxima of this potential, corresponding to minima in the interelectronic distance, are identified as possible recollisions. To avoid registering artificial maxima, we keep only the largest electron-electron potential within each quarter-cycle of the laser field. %provided that it exceeds a threshold value. In the present work this threshold is set to $0.05$~a.u.

\begin{figure}[t]
	\centering
\includegraphics[width=\linewidth]{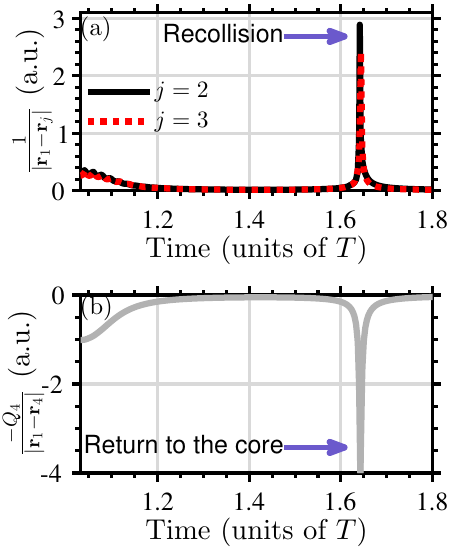}
\caption{Illustration of the recollision identification procedure. (a) Interelectronic potential energy for the electron pairs $(1,2)$ (solid black line) and $(1,3)$ (dotted red line) as a function of time. (b) Attractive electron-core potential energy between electron 1 and the core.}
\label{fig:schematic_rec}
\end{figure}

After identifying the maxima of the interelectronic potential energy, we verify that they correspond to actual recollisions by also identifying the times at which the quasifree electron makes its closest approach to the core. Each maximum of the interelectronic potential energy is associated with the closest-in-time return to the core. If such an association can be made, we define the time of the interelectronic potential maximum as the recollision time; otherwise, the possible recollision is discarded. Moreover, we note that electron-electron potential maxima involving different bound electrons but associated with the same return to the core are treated as the same recollision. These recollision events are then used to classify the direct and delayed ionization pathways, as described in Sec.~\ref{Section:Positive_mom_offset_pathway}. Further details of the recollision identification procedure and the identification of ionization pathways can be found in Ref.~\cite{Agapi3electron}.

In Fig.~\ref{fig:schematic_rec}, we present an example trajectory as an illustration of the procedure we follow to identify a recollision. Specifically, in Fig.~\ref{fig:schematic_rec}(a), we plot the interelectronic potential energy, $ \frac{1}{ | \mathbf{r}_i - \mathbf{r}_j | } $, as a function of time for the electron pairs $(1,2)$ (solid black line) and $(1,3)$ (dotted red line), starting from $t_0$ until shortly after the recollision time. We remind the reader that electron 1 is the initially tunnelling electron, while electrons 2 and 3 are the initially bound electrons. In Fig.~\ref{fig:schematic_rec}(b), we show the attractive electron-core potential energy, $ -\frac{Q_4}{ | \mathbf{r}_1 - \mathbf{r}_4 | } $, between electron 1 and the core.

As indicated by the purple arrow in Fig.~\ref{fig:schematic_rec}(a), a maximum in the interelectronic potential energy is identified for both electron pairs around $1.65T$. At approximately the same time, a minimum in the electron-core potential energy of the recolliding electron is observed, corresponding to its closest approach to the core. Therefore, the maxima in the interelectronic potential energy can be associated with a return of the quasifree electron to the core and is consequently identified as a recollision.

\bibliography{Bibliography}% Produces the bibliography via BibTeX.

\end{document}